# A Rapid-prototyping CMOS-RRAM Integration Strategy

Andreas Tsiamis[1*], Spyros Stathopoulos[1] and Themis Prodromakis[1]

[1] Centre for Electronics Frontiers, Institute for Integrated Micro and Nano Systems, School of Engineering, The University of Edinburgh, U.K.

* Corresponding Author: Andreas Tsiamis (Email: a.tsiamis@ed.ac.uk)

## Abstract

Moore's law has long served the semiconductor industry as the driving force for producing ever-advancing electronics technologies. However, given the economic implications and technological challenges associated with the present semiconductor scaling constraints, a shift from a traditional more Moore approach to a beyond Moore paradigm is desirable for sustaining the current pace of innovation beyond the established development route. Resistive random-access memories (RRAM) are one such beyond Moore technology that offers many avenues for innovation, and when integrated with mature complementary metal oxide semiconductors (CMOS), can extend CMOS capabilities in a scalable and power-efficient manner, both in terms of memory and computation. Nevertheless, as emerging and established technologies fuse, existing semiconductor-optimised manufacturing faces significant challenges, while the methodologies and complexities of integration are often not highlighted in depth, or overlooked at the expense of demonstrating the application-specific integrated-technologies. In this article, we focus on the integration, and detail a cost-effective, rapid-prototyping, and technology agnostic CMOS-RRAM integration strategy that employs hybridised wafer-level and multi-reticle processing techniques, supported by a systematic increased complexity approach. Leveraging the fact that CMOS technologies can be readily realised by taking advantage of mature front-end-of-line fabrication processes offered by semiconductor foundries, we establish an in-house RRAM development program that allows to combine fundamental material and device-level knowledge with custom-designed CMOS electronics. This approach utilises fully CMOS-compatible and transferable processes, ultimately enabling a seamless transition from research and development to volume production.

## Introduction

Since the dawn of monolithic integrated circuits (ICs) and driven by Moore's law[1], advances in semiconductor manufacturing have elevated complementary metal oxide semiconductors (CMOS) to a scalable, low-cost and robust platform for the development of a diversity of electronics technologies. Fuelled by innovations shaping the semiconductor nanofabrication landscape, more Moore has thus long served the industry as the driving force for producing ever-advancing CMOS technologies. Since its inception, Moore's law has been prolonged by a series of seminal engineering breakthroughs[2,3,4,5,6], the most recent being extreme ultraviolet lithography[7], which has emerged as the dominant technology for advanced semiconductor manufacturing, particularly for the 7 nm node and beyond. Utilising sophisticated light sources operating at high vacuum, next-generation high numerical aperture optics[8] and complex computational models that optimise the photolithographic process[9], this advanced technology is not only exceptionally intricate but is additionally driven by stringent requirements such as ultra-flat light collecting mirrors and defect-free multilayer reflective photomasks. As such, Moore's law is approaching its physical limits, while the technologies that have been instrumental for pushing towards these limits come at a significant cost. Given the economic implications and technological challenges associated with the present semiconductor scaling constraints, a shift from a more to a beyond Moore focus may prove catalytic for sustaining the current pace of innovation beyond the traditional development route.

Beyond-CMOS offers many avenues for innovation, enhancing electronics by harnessing emerging and disruptive technologies that employ novel materials, processes, devices, and architectures[10,11]. One such technology domain encompasses novel memory concepts, which not only have attracted considerable scientific attention, but are now gaining traction in industry with several being offered by semiconductor foundries, albeit as embedded solutions and for niche applications[12]. At the forefront of this rapidly evolving field are phase-change (PCM)[13], magnetic (MRAM)[14], ferroelectric (FeRAM)[15] and resistive random-access memories (RRAM)[16]. Often referred to as memristors[17], they are recognised for enabling a versatile application spectrum spanning non-volatile storage[18], reconfigurable electronics[19], in-memory and edge computing[20,21], and neuromorphic and bio-inspired systems[22,23].

A nascent technology that laid dormant at first[24], memristors are non-linear dynamic electronic devices whose inherent characteristics have long been observed and documented[25]. With the advent of the first such physical nanoscale devices, experimental validation[26] and theoretical framework[17] converged to initiate a period of intensified research and technological advancement. Groundwork on fundamental device physics, materials exploration and foundational architectures paved the way for the development and realisation of sophisticated, energy-efficient memristor-based systems that have been monolithically integrated with mature CMOS technologies[27,28] to extend their capabilities, both in terms of memory and computation. Such novel integrated solid-state systems are often conceptualised in scholarly work from an application-focused perspective that demonstrates their potential for expanded functionality and enhanced performance, while the description of their fabrication process and integration route is typically limited to key technical details[29,30,31,32]. However, as emerging and established technologies fuse, the cost of research and development (R&D) rises and the existing semiconductor-optimised manufacturing ecosystem is actively challenged by material and process incompatibilities, integration and scaling complexities, and device-level uncertainties. On account of this, the authors of this article recognise that systematic documentation of the integration methodologies (the journey), holds equivalent significance as the demonstration of the integrated technologies (the destination).

We select RRAM as the demonstrator integrated technology, owing to its competitive advantages that offer low-power consumption, fast switching with long retention characteristics, multi-state operation, high scalability and CMOS back-end-of-line (BEOL) compatibility[33,34]. Leveraging on the fact that CMOS technologies can be readily realised by taking advantage of mature front-end-of-line fabrication processes offered by leading semiconductor foundries, we establish an in-house RRAM development program that allows to combine fundamental material and device-level knowledge with custom-designed CMOS electronics. In particular, in this article we detail a cost-effective, rapid-prototyping, and technology agnostic CMOS-RRAM integration strategy that employs hybridised wafer-level and multi-reticle processing techniques, with a systematic increased complexity approach that is guided by earlier IC design and RRAM integration development cycles.

## Results and Discussion

### Development of CMOS Integrable RRAM Technologies

A CMOS integrable RRAM technology should fulfil all CMOS bound requirements, in addition to being able to demonstrate any desirable technology driven specifications. A key requirement is that the devices are fabricated using CMOS compatible and transferable processes, enabling to transition from R&D to volume production. Additionally, the programming and electroforming voltages of the devices should be below the supply voltages of the target CMOS node. Device specifications should be drawn from the merits of the technology, while the process of deriving the specifications, should be constrained to the CMOS bound requirements.

In readiness for the full integration cycle, the initial development of our RRAM technology is performed on 150 mm (thermally oxidised) silicon wafers. This is highly beneficial for early-stage development purposes, where material-based rapid-prototyping and technology characterisation performed directly on fully functional CMOS wafers could be cost-prohibitive and potentially hindered by the complexities of the integration process. The design follows a two-terminal metal-insulator-metal architecture, and the fabrication process uses standard contact photolithography for pattern transfer, while metallisation (bottom and top electrode (BE/TE)) and active (resistive switching) layer deposition are performed by e-beam evaporation, sputtering (including reactive) and atomic layer deposition (ALD). All device layers are defined either by photoresist lift-off or reactive ion etching (RIE). An optical microscopy image of a standalone memristor and its corresponding cross-sectional schematic (dotted line through device) can be seen in Fig. 1a. The memristors are laid out as 32 standalone devices per chip (3 x 3 mm) or alternatively as 32 x 32 crossbar arrays as shown in Fig. 1b, with each chip comprising devices of a single dimension. The selected active area device dimensions range from 60 x 60 $\mu m^2$ down to 1 $\mu m^2$, with the smallest being a representative critical dimension (CD) of CMOS integrated RRAM. The chip layouts are repeated across the wafer, allowing for high-throughput, full-wafer (1000's of devices) characterisation, when measured with a semi-automated probe system that is interfaced with an Arc ONE instrumentation platform[35].

We have previously utilised a variety of materials and deposition techniques for fabricating and characterising metal-oxide memristive devices, aiming to further understand the mechanisms that support resistive switching and ultimately to optimise the performance and enhance the reliability of the technology. Previous work spanned from identifying the conduction mechanisms of the well-received Pt/$TiO_{2-x}$/Pt stack[36], to exploring active layers such as $SnO_x$[37] that have received less attention, in addition to demonstrating the benefits of bilayer configurations such as $Al_xO_y/TiO_2$[33] and $HfO_2/Ta_2O_5$[38] and finally selecting appropriate electrode materials and identifying their interfacial characteristics, as they also hold a significant role for device performance[39]. Fig. 1c summarises our electrode and active layer material library and their deposition methods.

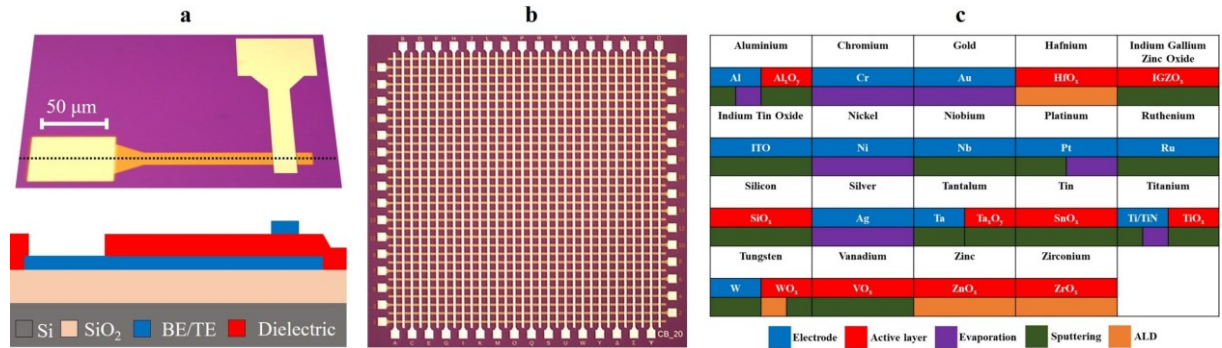

**Fig. 1: Memristive device architectures and associated library of device materials.** (a) Optical microscopy image and cross-sectional schematic of a standalone memristor (dotted line through device). (b) Optical image of a 32 x 32 crossbar array, 3 x 3 mm RRAM chip. (c) Electrode and active layer material library and their deposition methods.

A technique that we have previously explored[40] and has received attention as a means of improving the electrical characteristics of memristors, is the introduction of dopants in the active layer. The technique is applicable for lowering the electroforming voltages, which not only minimises the risk of irreversibly damaging the devices during the forming process, but crucially could allow devices to operate within the voltage limitations of the CMOS technology. For the development of an integrable RRAM we have selected nitrogen as the dopant, while considering the well-established and characterised titania[26,41] and hafnia[42,43] films as reference. $TiO_xN_y$ films were deposited using DC magnetron sputtering from a 99.995% titanium target either in an Ar rich plasma with $N_2$ and $O_2$ reactants or in a $N_2$ rich plasma with the addition of $O_2$ as co-reactant. The film that displayed the most suitable active layer characteristics was sputtered in an $N_2$ rich plasma at 500 W power, 1mT process pressure and $N_2/O_2$ gas flows of 20/5 sccm respectively (0.02 nm/s deposition rate). The active layer (25 nm thickness) was interfaced with Pt electrodes and the fabricated devices required both lower forming (~ 6.5 V) and switching voltages (~1 V) in comparison to their $TiO_x$ counterparts (~9V and ~1.5 V respectively). The $HfO_xN_y$ films were synthesised by thermal ALD using tetrakis dimethylamido hafnium IV (TDMAHf) as the hafnium source, $N_2$ plasma as the nitrogen source and either $H_2O$ or $O_2$ plasma as the oxygen source. The film with the optimal active layer characteristics was synthesised at 250 °C, using TDMAHf, $H_2O$ and $N_2$ plasma (5 sccm $N_2$ gas flow) at 300 W. In comparison to any titanium- or hafnium-based oxide and oxynitride variants explored, the devices fabricated using this $HfO_xN_y$ composition as an active layer (10 nm thickness) when interfaced with TiN electrodes, were found to display the most favourable electrical characteristics for a CMOS integrable RRAM. This included low forming (~2–4 V) and switching voltages (~1–3 V), with gradually quasi-analogue tuneable resistance.

**CMOS Electronics and Wafer-level Post-processing**

Several IC designs were taped-out based on a systematic increased complexity approach that followed the development cycles of the integration processes. Initial designs focused on simple test structures for process control and small memory arrays, followed by efforts on array up-scaling and finally complex array designs and new concepts based on RRAM technologies such as sensors, neural networks and adiabatic memories. Custom CMOS electronics were sourced from a silicon foundry and fabricated on 200 mm wafers (Fig. 2a), using a 1P5M 180 nm process (1.8/5 V supply voltages), with the option to end the fabrication process at metal-4, followed by deposition of, what would normally be, an interlayer dielectric (ILD) and finally planarisation.

The integration methodology detailed in this article is applied to the metal-4 version of the CMOS wafers, where the ILD-4 stack acts as the passivation layer and comprises ~650 nm thick $SiO_2$, capped by ~200 nm thick $Si_3N_4$. These materials and thicknesses are considered typical across the discussed technology nodes and prior to the industry shifting to low-k carbon-doped $SiO_2$ based dielectrics for the 90 nm node and below. To minimise the integration complexity, particularly for high-density architectures, the foundry passivation was then thinned down to ~100 nm by a means of etching and re-planarisation. First, a foundry-spun protective photoresist layer was removed using N-Methyl-2-pyrrolidone (NMP) at 65 °C, followed by a 30 min $O_2$ plasma at 350 W and 800 mT process pressure. The wafers were then blanket etched using 5% hydrofluoric acid (HF) with an observed 1:2

Si$_3$N$_4$/SiO$_2$ etch selectivity, that reduced the passivation thickness to ~ 200 nm SiO$_2$. It would be anticipated that this wet etch method would result in a thinned passivation with no significant local non-uniformities. However, the etched-surface topology was modified significantly matching that of the underlying metallisation pattern as can be seen from the profilometry scan and the corresponding optical image in Fig. 2b. While the cause of this effect is unclear, it may be the result of some form of local metal migration in the dielectric, which is likely to be dominant in the proximity of the two-layer interface. Therefore, although the etch may initially be uniform, as the passivation is progressively thinned, the etch rates for areas with and without underlying metallisation differentiate representing a local difference in material composition. Chemical mechanical polishing (CMP) was then employed to further thin the passivation and re-planarise the wafer surface. CMP was performed with a colloidal silica slurry, while the polishing pad was re-dressed at regular intervals. Platen and polishing spindle speeds were set to 35 and 30 rpm respectively, while the downforce pressure was set to 90 kPa. Post-CMP wafer cleaning was done in a 0.4% trimethylanilinium hydroxide (TMAH) solution with ultrasonics, followed by ultrasonics assisted soaking in de-ionised (DI) water at 65 °C, spin drying and finally O$_2$ plasma clean-up. Fig. 2c shows an atomic force microscopy (AFM) scan of a small section of the re-planarised wafer surface with a calculated mean roughness (S$_a$) of 0.21 nm and root mean square roughness (S$_q$) of 0.31 nm. The matched layout of the underlying metal layer is also shown, confirming a significantly reduced topology. Both reduced surface topology and roughness are favourable characteristics of the thinned CMOS passivation, with the first primarily minimising the photolithographic complexities of the RRAM integration process and the latter minimising any possible influence to the conduction mechanisms of the integrated devices.

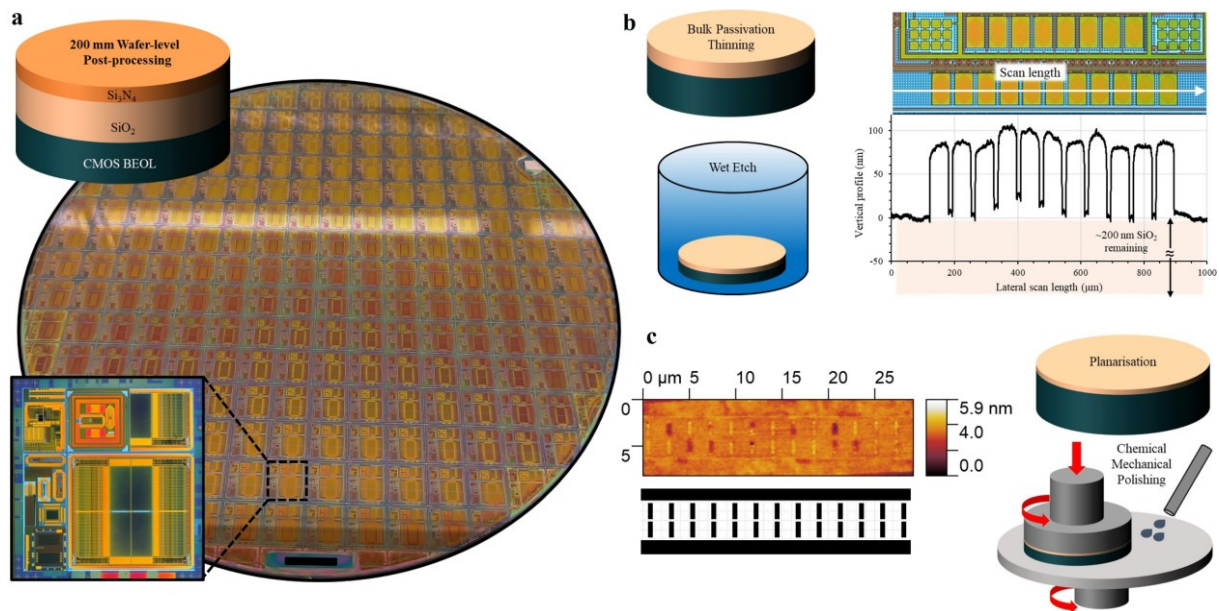

**Fig. 2: Wafer-level post-processing of memristors.** (a) Photograph of 200 mm foundry fabricated CMOS wafer. Inset shows an optical microscopy image of a multi-design reticle. (b) Foundry passivation is initially thinned by blanket wet etching. The representative profilometry scan (and corresponding optical image) of a section of the partially etched CMOS passivation reveals an etching-induced, topology-heavy wafer surface. (c) Final passivation thinning and re-planarisation by CMP. A representative AFM scan of a section of the re-planarised wafer surface, with the matching layout of the underlying metal layer.

We have previously demonstrated a methodology for adding functionality to foundry produced CMOS at singulated chip level[44]. Wafer-level post-processing, combined with stepper photolithography positions itsself at the opposite end of the integration spectrum. Both options are defined by merits and drawbacks, that are tied to cost of operation, complexity of processing technologies and speed of development cycles. For the work presented in this article, we have taken an approach that meets these options in the middle. Therefore, post passivation thinning, the CMOS wafers are diced into larger sized chips comprising the full reticle replicated several times. Although process equipment are normally set up and optimised for full wafer processing, the selected chip size offers adequate processing compatibility (no mounting on carrier substrates is required) and when combined with direct-write photolithography or e-beam lithography is ideally suited for enabling a fast development route towards a CMOS integrated technology, where both integration processes and integrated devices can be characterised using a reasonable number of nominally identical samples. This is supported by the proposed lithographic options that allow for straightforward modification of designs should layout patterns need to be revised, while their slow exposure speed (particularly at high resolution) is practically not applicable when patterning at the proposed scale.

**Multi-reticle CMOS-RRAM Integration**

*Integration Development: Small-scale Arrays*

Early-stage integration focused on a small array design, where RRAM devices were integrated to the CMOS BEOL to realise a 16 x 16 one-transistor-one-resistor (1T1R) crossbar array. With no added complexities arising from standard CMOS electronics, the selected design is ideally suited for the development and characterisation of the core integration processes, while process compatibility with CMOS can be verified with an electronically functional integrated array. A summary of the proposed RRAM integration process is schematically presented in Fig. 3. Briefly, the foundry passivation (Fig. 3a) is initially selectively etched to expose the top CMOS metallisation (Fig. 3c). This is followed by patterning of the RRAM stack (Fig. 3e, 3g and 3i) and finally the etched trenches are metal-filled to ensure good connectivity between RRAM and CMOS BEOL (Fig. 3k). The adaptation of this pseudo-via approach eliminates the need for establishing damascene and metal CMP during early-stage development, while these wafer-level-only processes can be incorporated at a latter phase, when integration transitions to volume fabrication lines. Fig. 3b shows an optical image of a section of the foundry produced CMOS array. Each column in the array comprises a wide metal-4 track serving as a connection placeholder for the TE of all RRAM devices in a column (bit-line), while each row consists of small metal islands that are interconnected to the drain of pMOS FETs and also serve as individual connection placeholders for the BE of each RRAM in a row.

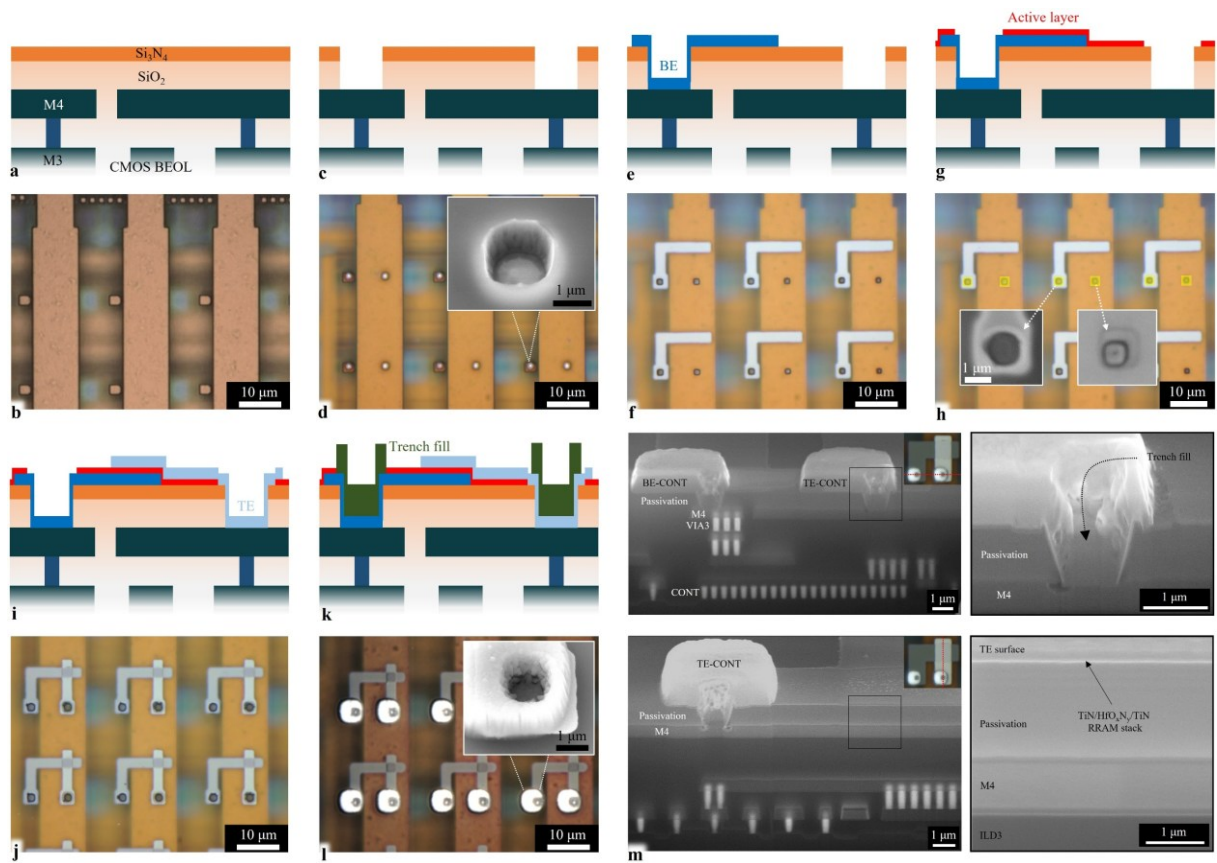

**Fig. 3: CMOS-RRAM integration and physical validation.** (a) Cross-sectional schematic of foundry produced CMOS BEOL. (b) Optical microscopy image of foundry fabricated CMOS array (part of). (c) Selective passivation etch (vias). (d) Optical image of the passivation openings (inset shows SEM imaging of trench profile). (e) RRAM BE deposition and patterning. (f) Optical image of 2 μm TiN BEs. (g) RRAM active layer deposition and patterning. (h) Optical image (post dielectric deposition) with overlayed mask (yellow) of active layer via-etch (insets show UV images of the re-opened vias) (i) RRAM TE deposition and patterning. (j) 16 x 16 array (part of) with 2 x 2 μm$^2$ TiN/HfO$_x$N$_y$/TiN RRAM devices integrated with CMOS (1T1R). (k) Trench metal-fill.

(l) Optical image of 1T1R devices with filled pseudo-vias ensuring good connectivity between CMOS and RRAM (inset shows SEM imaging of a single fully filled trench). (m) SEM imaging (52° stage tilt) of FIB-milled CMOS-RRAM cross-sections at the CMOS-RRAM interconnect interface (top) and at the device interface (bottom).

Prior to post-processing, all previously diced CMOS chips underwent an extended $O_2$ plasma (30 min) that dehydrated their surface and removed organic residue. To create RRAM connectivity openings to the top CMOS metal, the chips were first treated with hexamethyldisilazane (HMDS) vapour to enhance photoresist adhesion and then spin-coated at 4000 rpm with positive photoresist, and soft baked (SB) at 115 °C for 90 s. The resulting ~2.3 μm photoresist thickness was optimised for resolving the smallest dimensions of the desired photolithographic pattern, while additionally maintaining integrity as a protective mask through a prolonged etch process. The etch pattern (L1) comprising isolated 1.5 μm wide squares was exposed by optical direct-write lithography (DWL) using a 405 nm light source through a 20X apochromatic objective lens, with an exposure dose of 150 mJ cm$^{-2}$ and a focal offset of +2.5 μm. Pattern exposure was repeated across all reticles of the diced chips. Layer to layer alignment can be performed globally at multi-reticle level, or locally per individual reticle. This is guided primarily by the alignment tolerance of the lithographic process and the alignment accuracy of the exposure tool. This was followed by a post-exposure bake (PEB) at 115 °C (90 s), photoresist development in 0.26N developer (60 s) and a hard bake (HB) at 115 °C (90 s). It should be noted, that due to the complexity added by the presence of the underlying CMOS BEOL multilayer interconnect stack, all photolithographic processes were first optimised for the desirable pattern fidelity using focus-exposure matrices with the matching post-processing layouts. The chips for this work were sourced from wafers where the passivation was not thinned as previously described. This was only the case when integrating low-density designs such as the one detailed here, where a deep trench surface topology did not affect any subsequent photolithographic steps and processes. Therefore, the full $Si_3N_4$/$SiO_2$ stack was etched by RIE in a $CHF_3$/Ar (20/20 sccm gas flow) plasma at 100 W power and 25 mT pressure, to form ~850 nm deep openings. At the expense of a slower etch rate this process produced features with near-vertical sidewalls, exhibited a favourable (~ 1:1) dielectric/photoresist selectivity and resulted in no noticeable photoresist reticulation. Finally, the remaining photoresist mask was removed using NMP at 65 °C and the chips were rinsed with isopropyl alcohol (IPA). Fig. 3d shows an optical microscopy image of the passivation openings, where the bright white features correspond to the exposed top CMOS metal. A scanning electron microscopy (SEM) image (inset) of one the openings captured at a 30° tilt angle is also presented.

To pattern the BE of the RRAM devices, the chips were first spin-coated at 5000 rpm with image reversal photoresist, followed by a SB at 105 °C (60 s) to produce a 500 nm thick photoresist layer. It should be noted that all photoresist coating steps included preceding sample dehydration and HDMS treatment steps. The BE pattern (L2) comprising right cornered 1, 2 or 5 μm wide tracks was exposed by optical DWL with an exposure dose of 25 mJ cm$^{-2}$. No other previously detailed DWL parameters were altered. This was followed by an image reversal process consisting of a PEB at 120 °C (120 s) and a blanket ultraviolet (UV) exposure, and finally photoresist development (0.26N, 60 s) and a brief descum (60 s) by $O_2$ plasma. A 50 nm thick TiN layer was then deposited by DC sputtering from a 99.995% titanium target in a $N_2$ rich (20 sccm gas flow) plasma at 600 W power and 2 mT process pressure (0.06 nm/s deposition rate). Prior to deposition and without breaking vacuum, a brief 100 W Ar mill removed any surface oxide from the exposed CMOS metallisation that would subsequently be capped by the TiN layer. The electrodes were ultimately patterned by photoresist lift-off using NMP at 65 °C and the chips were rinsed with IPA. Fig. 5f shows an optical image of a small section of one of the arrays, in this case integrated with 2 μm wide BE features.

A 5nm thick $HfO_xN_y$ active layer was then blanket deposited on the chips by ALD as detailed previously for the optimal active layer, and the thickness of the film was determined by in-situ ellipsometry. To re-expose the metal openings (TiN capped BE contact and CMOS metal-4 TE contact placeholder) the chips were first spin-coated at 5000 rpm with 1 μm positive photoresist, followed by a SB at 90 °C (60 s), exposure by optical DWL with a dose of 80 mJ cm$^{-2}$, PEB at 110 °C (60 s) and photoresist development (0.26N, 60 s). The layout of the etch pattern (L3) comprising 2 μm wide openings is overlayed to the L1 openings as highlighted by the features in yellow in Fig. 3h. The $HfO_xN_y$ film was then etched by RIE using the process parameters detailed for the CMOS passivation etch. The remaining photoresist mask was then removed as described previously. The insets of Fig. 3h show magnified UV images of the re-opened connections, with the faint square patterns representing the edges of the newly etched features.

50 nm thick TiN TEs were then fabricated following the procedure detailed for the BE fabrication process. The TE pattern (L4) comprises vertically running 1, 2 or 5 μm wide tracks intersecting the BE pattern. Fig. 3j shows an optical image of a section of the 16 x 16 array with 2 x 2 μm$^2$ TiN/HfO$_x$N$_y$/TiN RRAM devices integrated with the CMOS in a 1T1R configuration. It should be noted that it is anticipated that a post-deposition native oxide will form on the surface of the TiN layer, effectively altering the interconnection conductivity of the electrode. To minimise this effect, the TE is normally capped with a thin layer (~20 nm) of Pt, which is DC sputtered, without breaking vacuum, from a 99.99% platinum target in an Ar rich (20 sccm gas flow) plasma at 400 W power and 3 mT process pressure (0.4 nm/s deposition rate).

To metal-fill the etched trenches, the chips were first spin-coated at 3500 rpm with 3 μm negative photoresist followed by a SB at 105 °C (60 s), exposure by optical DWL using a 365 nm light source and a dose of 155 mJ cm$^{-2}$, PEB at 105 °C (60 s), photoresist development (0.26N, 90 s) and a brief descum (60 s) by O$_2$ plasma. The layout of the fill pattern (L5) comprising 4 μm wide squares was overlayed to the L1/3 openings. A 10/850 nm Ti/Al bi-layer with Ti serving as adhesion, was then deposited by e-beam evaporation (0.2/0.6 nm/s deposition rate) at 10$^{-7}$ Torr (or lower) pressure and finally photoresist lift-off defined the fill pattern. Aluminium was selected as the fill metal, as it is frequently utilised for CMOS interconnect and vias, particularly at the technology node discussed in this article. Copper and tungsten, which offer lower resistivity and enhanced resistance to electromigration respectively, are compatible options for integration at the 130 nm node and beyond. Fig. 3l shows an optical image of the integrated RRAM devices with their CMOS connecting trenches fully filled, while the inset shows a SEM image of a single filled trench. The metal protruding from the chip surface is the section of the pattern extending past the trenches.

To cross-sectionally validate the integration process, focused ion beam (FIB) milling was performed along the CMOS-RRAM interconnect and device interfaces, as can been seen from SEM imaging in Fig. 3m. The integration process appears to have not caused any physical damage to the underlying CMOS BEOL-stack, while the CMOS passivation trenches are homogenously filled with the thick aluminium, and the TiN layer is conformally coated on the sidewalls of the trenches, in a manner similar to a barrier layer. Interestingly enough, the CMOS passivation seems to have been over-etched, and the RIE process has clearly partially attacked the top CMOS metallisation, thus modifying the local topology. However, the etched metal has mostly been recovered through the aluminium deposition, in which case this effect is unlikely to be of any significance. Similarly, SEM imaging through the device cross-section shows that the patterned RRAM stack appears to be homogeneous.

To enable access for wire-bonding or direct electrical probing of the integrated array, a final process was utilised to create openings to the foundry passivated pads. First the chips were spin-coated at 2000 rpm with 3 μm positive photoresist, followed by a SB at 115 °C (90 s), exposure by optical DWL using a 405 nm light source through a 5X achromatic objective lens with a dose of 200 mJ cm$^{-2}$, a PEB at 115 °C (90 s) and photoresist development (0.26N, 60 s). The layout of the pad etch pattern (L6) comprises 57 x 67 μm rectangles, slightly smaller than the foundry designed metal pads. The film was then etched by RIE using the previously detailed passivation etch, with the addition of a brief CF$_4$/O$_2$ finishing etch (60/4 sccm gas flow) at 150 W and 60 mT pressure, which minimises the risk of damaging the metal layer during the pad exposure process. Fig. 4a shows an optical image of a standard multi-reticle (3 x 2) chip used in the integration process. The expanded view shows a 1T1R integrated memory array chip with exposed pads for the word-lines, bit-lines, source-lines and FET bulk connections.

The integration process was electronically validated by investigating the forming and switching characteristics of sample 1T1R arrays, using a custom-designed measurement platform (see methods). The basis of the array used as a benchmark platform is configured such that a single gate-driven line (word-line) controls a column of devices and a single source-line controls the biasing of the pMOS that form the core of the circuit. Current read-out is done from the free electrode (TE), ie. the end of the RRAM not directly connected to the pMOS (bit-line). The transistors are 5 V devices and are active low. The voltage supplied to the RRAM is determined by the voltage drop between the drain ($V_d$: 5–0 V) and the free biasing electrode ($V_R$: -5–5 V) minus the voltage drop across the transistor ($V_{ds}$), i.e., $\Delta V_R = V_d - V_R - V_{ds}$. Fig. 4b shows a resistance map for a 16 x 16 1T1R array, where all the cells are initially measured (read at 0.5 V) at their pristine state (MΩ range). Fig. 4c presents an optical image of a fabricated array where the highlighted devices have undergone an electroforming procedure that conditions them to within their operating resistive range (10–100 kΩ, read at 0.5 V), as can be seen from the resistance map of Fig. 4d (diagonal selection). This consequently allows to manipulate the devices freely, and in this particular case it is typically achieved by applying a set of a 1000 1 ms-wide pulses at voltages ranging from 2 to 3 V although devices can also be formed by applying an I-V curve as shown in Fig. 4e. Similarly, devices can be programmed to different resistive values with either 100 μs pulses ranging from 1–3 V using a scheme similar to[33] or directly using I–V sweeps as shown in Fig. 4f.

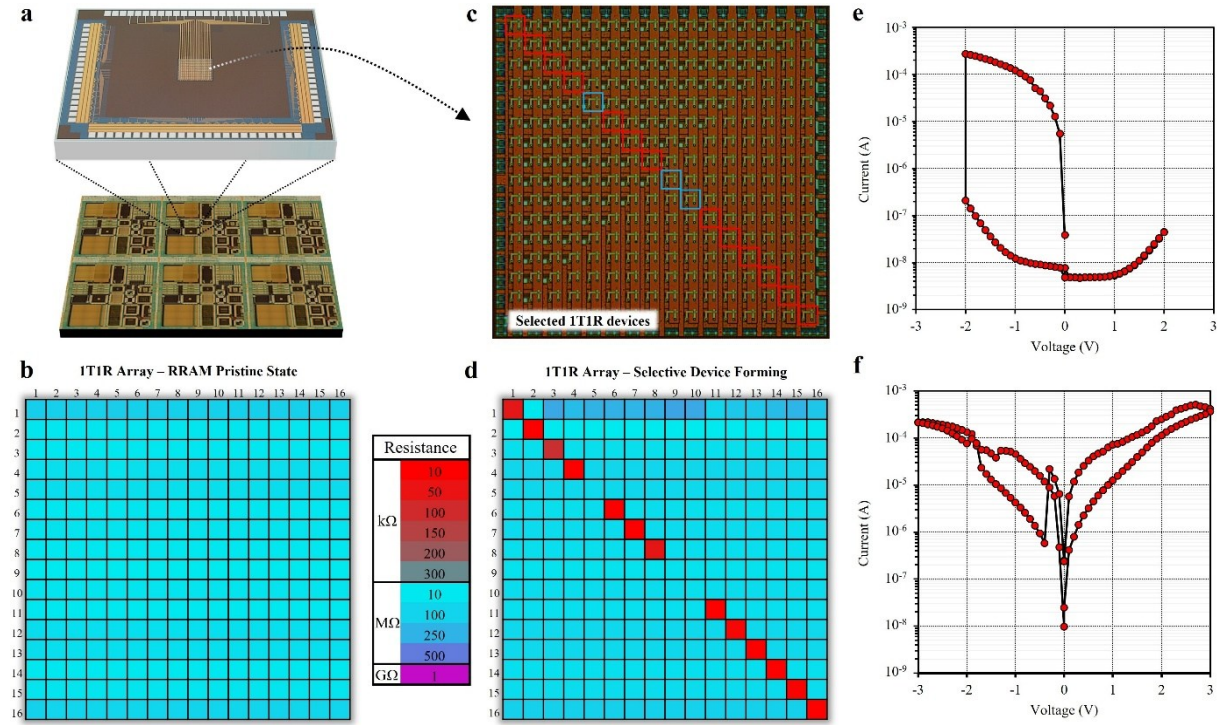

**Fig. 4: Electronic validation of CMOS-RRAM integration process.** (a) Typical integrated multi-reticle chip. Expanded view shows a single fully-integrated CMOS-RRAM die. (b) Resistance map of array with devices at their pristine state. (c) Optical image of 16 x 16 1T1R (pMOS-TiN/HfO$_x$N$_y$/TiN) array with highlighted devices selected for electrical characterisation. (d) Resistance map of array after one electroforming cycle, with most of the (diagonally selected) devices at their operating resistive range. (e) Representative RRAM forming (1T1R cell (12,12)) and (f) switching (1T1R cell (3,3)) I-V characteristic curves.

It should be noted that while this representative electrical characterisation validates the integration strategy and confirms process and technology compatibility, it does not examine integration uniformity and repeatability, and their potential influence to device characteristics. The example resistance map of these formed RRAM cells (Fig. 4c) shows that some of the devices remain at their pristine state. Although unlikely, this could be attributed to local dimensional, and thin-film thickness and roughness non-uniformities arising by patterning and process-induced effects that would directly impact device-to-device characteristics. More likely, these variabilities are related to the stochasticity in RRAM filament formation. Typical process-induced variations are likely to be systematic, and would not normally be so localised. Such effects can be quantified and potentially correlated to device characteristics, initially at multi-reticle level, at the prototyping stage, and ultimately as standard wafer-level uniformity and repeatability measurements for process control.

*Array Up-scaling*

While the core processes of the monolithic integration were developed using small, low-density array designs, they are largely transferrable to more complex, large, and high-density memory arrays. Nonetheless, this transition is contingent on several processing factors that should be taken into consideration. For this purpose, we have designed a CMOS-based, RRAM characterisation platform with on-chip programming and read capabilities, comprising 4 identical and independently controlled 512 x 512 sub-arrays that when integrated with RRAM have the capacity to characterise up to 1 million cells[45] in a 1T1R configuration. An optical microscopy image of the architecture can be seen in Fig. 5a. Supplementary Fig. 1a shows a layout snippet of a dense 1 μm RRAM electrode configuration, with a 2:1 line-to-space feature ratio through the topology-heavy, CMOS-RRAM interface (dotted line), designed for optical DWL-based processing of the 1Mbit array. The significance of thinning the foundry passivation is evident by comparing the cross-sectional schematics of Supplementary Fig. 1a, which illustrate the desired electrode photolithographic patterns (photoresist profile), post metallisation, defined both on ~850 nm, deep trench and ~100 nm, shallow trench topologies. Clearly, a deep trench topology is unlikely to support defect-free, high-density patterning, as it is impacted by factors such as locally non-uniform photoresist surface coverage and significant local spatial variations in focal length on the exposure field (beyond the depth of focus). This is compounded by the fact that narrow, deep trench passivation topologies cannot be realised using thin, high-resolution photoresist masks alone, and would require the use of supplementary hard mask layers to enable pattern transfer, thereby introducing added complexity to the integration process. In comparison, a deep trench topology presents little challenge for low-density array patterning as illustrated in Supplementary Fig. 1b, and previously demonstrated.

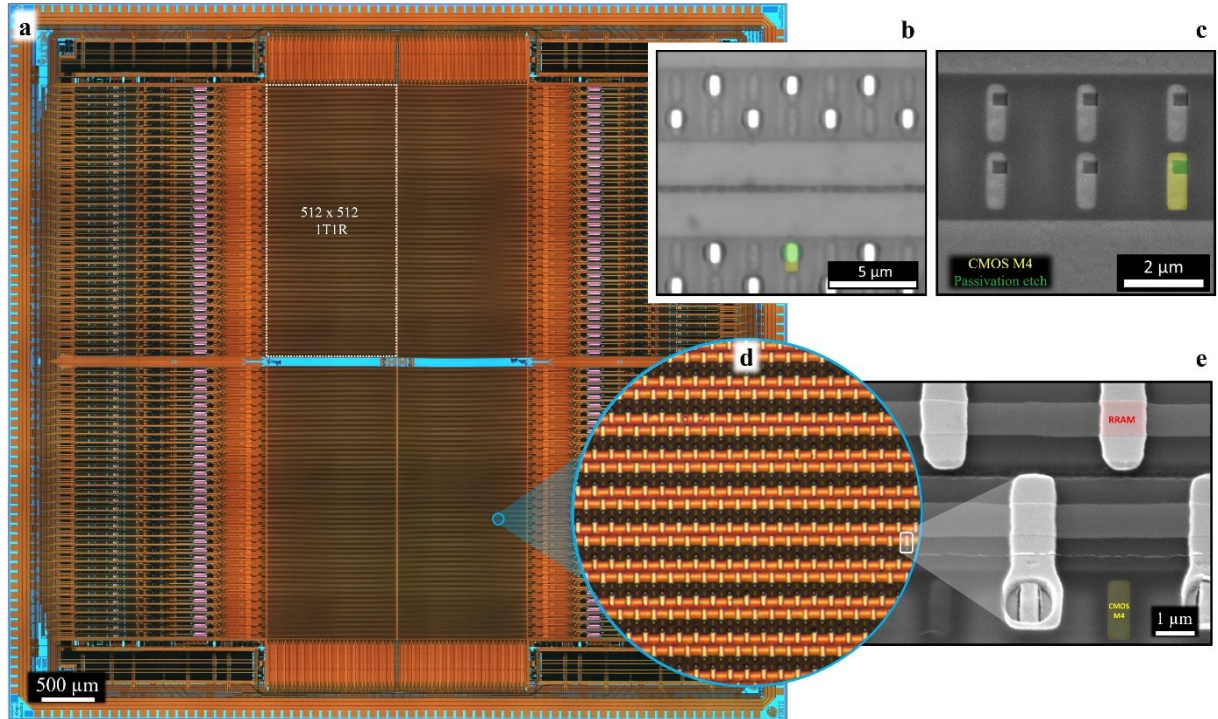

**Fig. 5: CMOS-RRAM array up-scaling**. (a) Optical microscopy imaging of a 1Mbit 1T1R characterisation platform comprising 4 identical 512 x 512 sub-arrays. (b, c) Comparison of passivation openings (by RIE) to expose the top CMOS metallisation, patterned by (b) optical DWL (UV imaging) and (c) e-beam lithography (SEM imaging). (d) Magnified optical imaging of a section of the integrated RRAM array (with relative size highlighted within the full array) and (e) high magnification SEM imaging (30° stage tilt) highlighting a single 1 µm$^2$ RRAM cell.

As the density of the RRAM array increases and the minimum CD decreases, the integration methodology is in part informed by the capabilities of the available lithographic technologies. Each of the 4 sub-arrays of the 1Mbit design, are laid out such that the top CMOS metal of each row in the sub-array comprises a long track serving as a connection placeholder for a common BE across all RRAM cells in the row, while each column consists of small metal islands that serve as individual connection placeholders for the TE of each RRAM cell in the column. The minimum relevant CD is 500 nm for the width of the metal-4 island, with a pitch of ~2 µm. Supplementary Fig. 2a and Fig. 2b show the layouts for producing CMOS-RRAM connectivity openings, designed both for optical and e-beam DWL respectively. While, the larger BE opening is laid out in the same manner for both lithographic options, this is not the case for the TE openings. The e-beam layout uses 400 nm wide openings, slightly smaller than the underlying metal, while optical patterning, which in comparison is limited in CD and pitch resolution, employs a sparse array layout with larger 1 µm openings that extend beyond the underlying CMOS metal. Both lithographic options were trialled for pattern transfer, with optical DWL following identical procedures and parameters for photoresist coating and exposure, as those detailed for the patterning of the active layer of the 16 x 16 array. For e-beam DWL, the chips were first dipped in a cationic priming agent to enhance resist adhesion, and then rinsed with DI water and baked at 180 °C (5 min). The chips were then spin-coated at 2000 rpm with ~300 nm negative high-contrast e-beam resist, followed by a SB at 150 °C (120 s). The etch pattern was then exposed using a 32 nA beam current with an exposure dose of 342 µC/cm2. This was followed by resist development in amyl acetate (60 s), rinsing with IPA, HB at 130 °C (60 s) and a brief descum by O$_2$ plasma (60 s, 150W). Multi-reticle substrates with the thinned SiO$_2$ passivation were finally etched by RIE using our standard CHF$_3$/Ar-based plasma process to form ~100 nm deep, and 400 nm or 1 µm wide openings. Fig. 5b and Fig. 5c show representative imaging of the passivation openings as defined by the two lithographic options (Supplementary Fig. 3b shows the larger openings that provide connectivity to the BE). Clearly, the optical DWL-based passivation etch extends beyond the width of the underlying CMOS metal, which to an extent, would normally act as an etch stop, and thus allow for relaxed over-etch process tolerances. Therefore, to minimise over-etching into the underlying ILD-3, the RIE process should be supported by end-point depth-profiling metrology. Supplementary Fig. 2c and Fig. 2d present the layouts for integrating the 3-layer RRAM stack, to realise a 0.5Mbit (sparse) 1T1R array with 1 $\mu m^2$ RRAM or a full 1Mbit 1T1R array with 500 x 500 $nm^2$ RRAM, using optical and e-beam DWL-based integration respectively. While e-beam DWL-based processing can support the integration of a full-density array, potentially allowing the fabrication of RRAM features with deep-nm dimensions, optical DWL-based processing is in comparison significantly restricted by the resolution limit of the optics, and is normally unable to resolve the sub-micrometre spacings required for the full-density array implementation. Nevertheless, optical DWL-based processing offers a cost-effective and rapid-prototyping route, which at early-stage development and when employed with resolution enhancement techniques, is often sufficient for demonstrating proof-of-concept (subset) integration of complex and dense array designs.

We demonstrate RRAM integration using our optional oxynitride-based (Pt/$TiO_xN_y$/Pt, 10/10/20 nm thickness) material stack, which in-part explores the utilisation of alternative layer deposition and processing techniques. Platinum RRAM electrodes were deposited by evaporation (0.06 nm/s deposition rate) at $10^{-7}$ Torr (or lower) pressure, while the $TiO_xN_y$ active layer was deposited by DC sputtering using process parameters that have been previously detailed. The BEs of the RRAM, which comprise long tracks spanning a full row of devices (common per row) were patterned by lift-off (see Supplementary Fig. 3b), using the methodology that has been previously described. The active layer was in this case also patterned by lift-off (see Supplementary Fig. 3c), owing to the fact that sputtering produces, in comparison, less conformal films than the highly-conformal ALD process. This, in turn, enables for defect-free (fencing artefacts) lift-off patterning, which would otherwise be difficult to attain, even when employing highly undercut photoresist patterns. At half-array density, the patterning process for the TE features of the RRAM remains challenging for this photolithographic approach. To produce the desired pattern we utilise double patterning, a well-established resolution enhancement technique that has been successfully implemented using several methodologies. In this case, by exploiting the fact that the alignment accuracy of the available photolithographic technology surpasses the minimum CD resolution, the half-array is fabricated using a two-step process, wherein the smallest feature size is realised as a cumulative outcome of both processes, rather than being resolved in a single step, as can been seen from the optical images of Supplementary Fig. 3c and Fig. 3d. Finally, Fig. 5d shows an optical image of a larger section of the integrated 0.5Mbit 1T1R array (on-chip circle represents the actual size relative to full chip), while high magnification SEM imaging in Fig. 5e highlights a single 1 $\mu m^2$ RRAM cell, with the trench revealing the interface between CMOS BEOL and RRAM. The CMOS metal-4 island protruding within the trench is an indication that in this case the passivation has been slightly over-etched into the ILD-3.

*Applications-focused CMOS-RRAM Architectures*

Ultimately, integrated RRAM technologies can coexist with and enhance a plethora of CMOS-based architectures, to realise systems that offer several pathways towards a beyond-Moore era. Although the challenges of RRAM integration are to some extent associated with the specific CMOS architecture being integrated with (i.e., developing RRAM technologies with targeted specifications), the proposed integration strategy and core fabrication processes are architecture-agnostic. The development and realisation of high-density arrays practically represents the peak in monolithic-integration complexity, which in turn, then provides the platform for transferring the same methodologies, technologies and processes to any proposed architecture. Example architectures are demonstrated in Fig. 6, which shows optical microscopy images of several application-focused RRAM-enabled CMOS chips, using TiN/$HfO_xN_y$/TiN (50/5/50 nm) RRAM with dimensions ranging from 1 to 2 x 2 $\mu m^2$. The RRAM cells were designed and populated in a manner that meets the specifications and BEOL layout of each architecture. These architectures include a 1T1R multi-function sensor interface for neuromorphic and bio-inspired analysis[46], a 9T4R analogue content addressable memory for analogue template matching at the edge[47], an analogue domain aggregator system for neural network accelerators[48], and 1T1R and 2T1R radiation-hardened (total ionisation dose and single-event effects) memory cells, to determine radiation tolerance within functional memory arrays, for use in high-radiation environments[49]. All architectures have been integrated with RRAM devices using the methodology and fabrication processes detailed in this article.

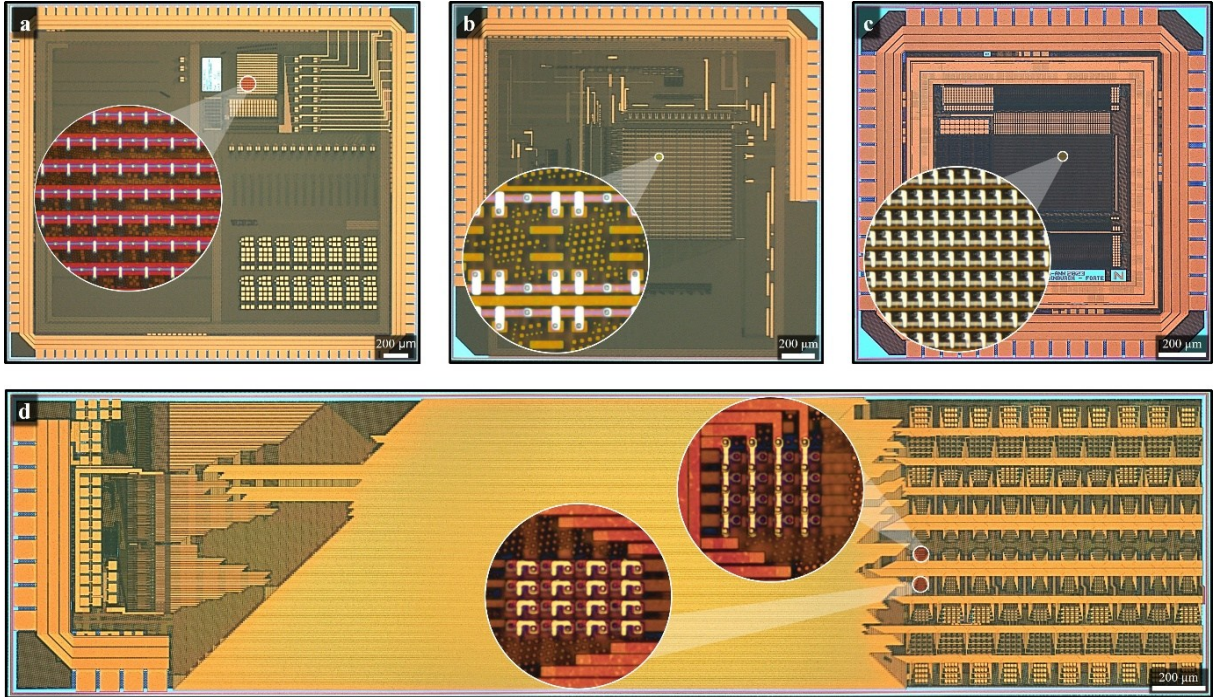

**Fig. 6: Application-focused CMOS-RRAM architectures.** Optical microscopy images of (a) 1T1R neural sensor interface[46]. (b) 9T4R analogue content addressable memory[47]. (c) 1T1R neural network accelerator[48]. (d) 1T1R and 2T1R radiation hardened memory cells[49].

## Conclusion

Beyond-CMOS technologies, such as emerging RRAM have become increasingly popular, offering wide-ranging opportunities to innovate alongside and beyond the established Moore scaling. In this article we have identified a rapid-prototyping integration strategy that combines wafer-level and multi-reticle post-processing techniques, to enhance CMOS electronics through integration with RRAM, in a technology-agnostic manner. RRAM prototypes, which are suitable for integration with CMOS, were first developed on standard silicon wafers, while early-stage post-processing of CMOS at wafer-level was utilised to realise a platform for integration. The core integration processes were first developed and validated on multi-reticle substrates using small-scale arrays comprising transistor-RRAM cells, which were then adapted and transferred to high-density arrays, and finally implemented on applications-specific architectures. The integration strategy utilises fully CMOS-compatible and transferable processes, allowing to transition from R&D to volume production, and in principle can be implemented with any memristive technology.

## Materials and Methods

The development and integration of the CMOS-RRAM technology was carried out in a class 10 (ISO4) cleanroom (21 °C, 40% relative humidity). CMP was performed with a Presi Mecapol E460 using Klebosol 30 HB 50 slurry. All wafers were diced with a DISCO DAD3350 dicing saw. Wafers and multi-reticle substrates were spin-coated with photoresist using a POLOS SPIN200i spinner. Several photoresist types were utilised. Positive photoresists, SPR 350 for thinner and SPR 220-3.0 for thicker coatings (> 2μm), were used for RIE processes. Similarly, AR-P 6200 (CSAR 62) e-beam resist was used for RIE. Image reversal AZ 5214E and negative AZ nLOF 2035 photoresists were used for lift-off. Microposit 1165 remover was used as photoresist solvent, while remover AR 600-71 was used as e-beam resist solvent. At the technology development stage, wafer-level RRAM patterning was performed with a Karl Suss MA8 Gen 2 Mask Aligner, while CMOS-integrated RRAM patterns were exposed at multi-reticle level with either a DMO MicroWriter ML3 Pro (optical DWL) or a RAITH EBPG (e-beam lithography). Positive photoresists were developed with MF26-A developer, while image reversal and negative photoresists were developed with AZ 726-MIF. Thin-films were deposited either with a Veeco Fiji Gen 2 ALD system or with Angstrom Engineering, Nebula sputtering and EvoVac e-beam evaporation systems. RIE and $O_2$ plasma processing was performed with a JLS RIE80 and an Electrotech 508 barrel asher, respectively. Optical microscopy images were captured with Leica DM12000 M and DM8000 M microscopes, and with a Reichert-Jung Polyvar-Met. FIB milling was performed with FEI Helios Plasma FIB-SEM, SEM imaging was carried out with a Tescan VEGA3, AFM surface analysis was conducted with a Park NX20 and profilometry scans were acquired from a Veeco Dektak 8 advanced development profiler. The thickness of ALD deposited thin-films was measured in situ with a Woolham iSE spectroscopic ellipsometer, while all other thin films were measured with either a Film Sense FS-1EX Gen3 ellipsometer or a Nanometrics Nanospec 3000 Reflectometer. The CMOS-RRAM integrated devices were electrically characterised with an ArC Instruments, ArC TWO tool[50], using a custom Python-scripted measurement interface and a custom interface board for on-chip measurements.

## Acknowledgements

This work was supported in part by the Engineering and Physical Sciences Research Council (EPSRC) Functional Oxide Reconfigurable Technologies (FORTE) Programme under Grant EP/R024642/2, in part by the EPSRC Low-Power, High-Speed, Adaptable Processing-In-Sensing Capability (ProSensing) Project under Grant EP/Y030176/1 and in part by the Royal Academy of Engineering (RAEng) Chair in Emerging Technologies under Grant CiET1819/2/93.

The authors thank Hannah Levene for assisting with AFM measurements and ALD deposition, Rahul Ramesh for assisting with ALD deposition, Camelia Dunare for assisting with wet processing, Graham Wood for assisting with wafer dicing, and Peter Lomax, Stewart Ramsay and Ewan McDonald for wider cleanroom support.

## Conflict of interest

The authors declare no competing interests.

## Author Contributions

A.T. developed the integration strategy, designed, fabricated and physically validated the integrated devices, assisted with device electrical characterisation and prepared the manuscript. S.S. developed the integration strategy, designed and assisted with the fabrication and physical validation of the integrated devices, electrically characterised the devices and assisted with manuscript preparation. T.P. supervised the project and provided critical guidance and feedback on manuscript preparation, the integration strategy, and device design, fabrication and characterisation. All authors reviewed and approved the final manuscript.

# A Rapid-prototyping CMOS-RRAM Integration Strategy

Andreas Tsiamis[1*], Spyros Stathopoulos[1] and Themis Prodromakis[1]

[1] Centre for Electronics Frontiers, Institute for Integrated Micro and Nano Systems, School of Engineering, The University of Edinburgh, U.K.

* Corresponding Author: Andreas Tsiamis (Email: a.tsiamis@ed.ac.uk)

**Supplementary Information**

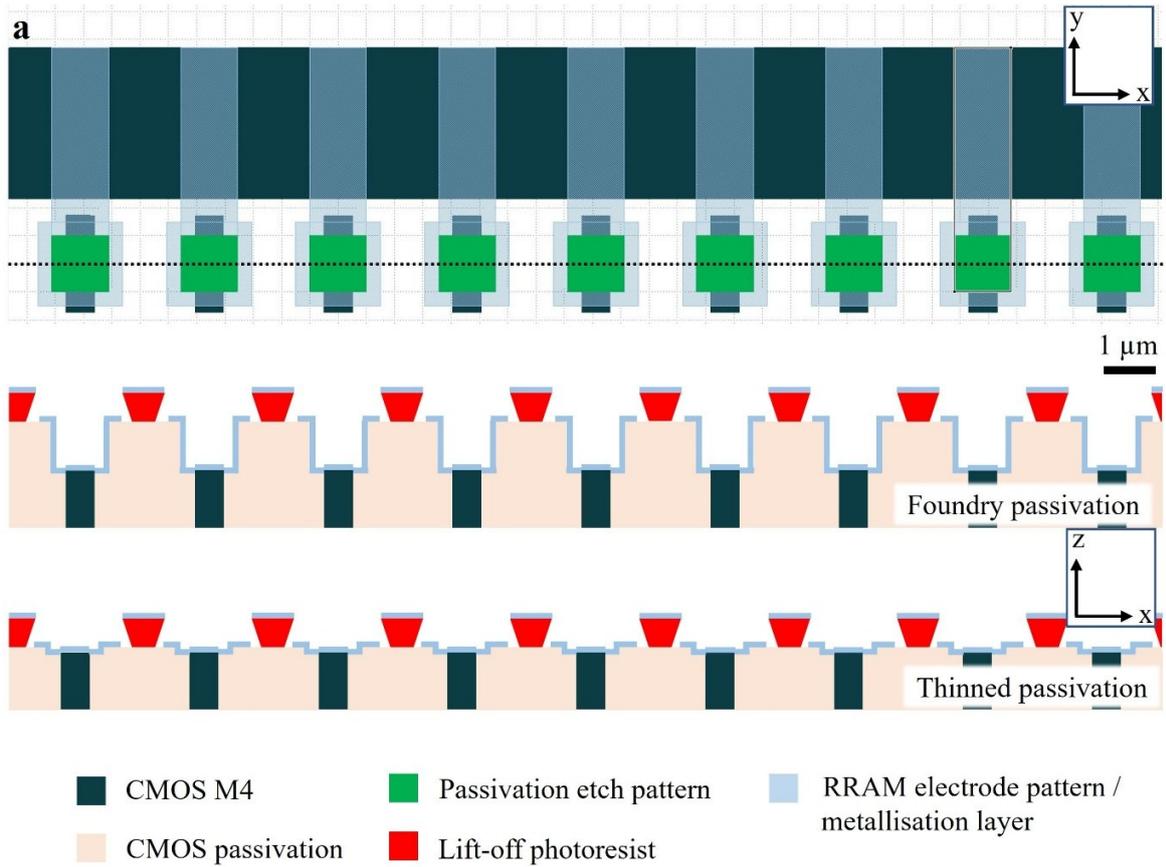

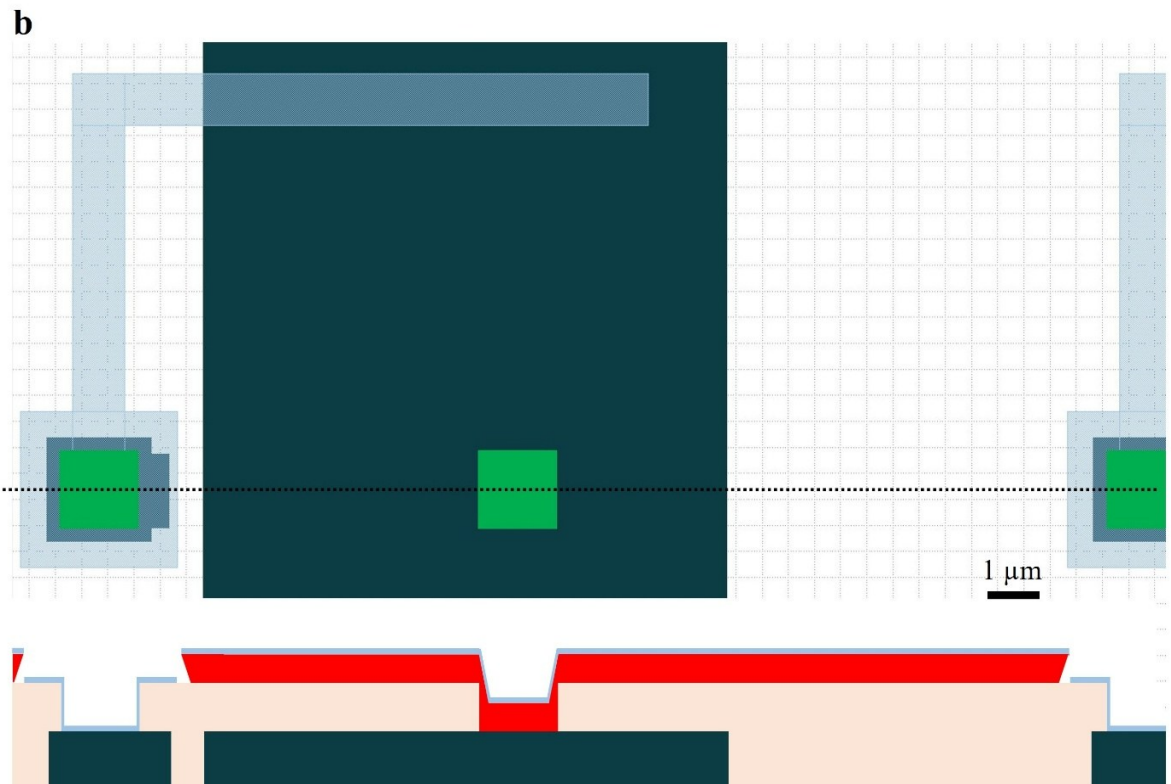

**Supplementary Fig. 1: Photolithographic patterning of high- and low-density arrays on deep and shallow surface topologies.** (a) Layout snippet of dense array (515 x 512 1T1R) patterning (top) and cross-sectional schematics of desired post-metallisation patterning on deep (middle) and shallow (bottom) trench topologies. (b) Layout snippet of low-density (16 x 16 1T1R) array patterning (top) and cross-sectional schematic of desired post-metallisation patterning on deep (bottom) trench topology.

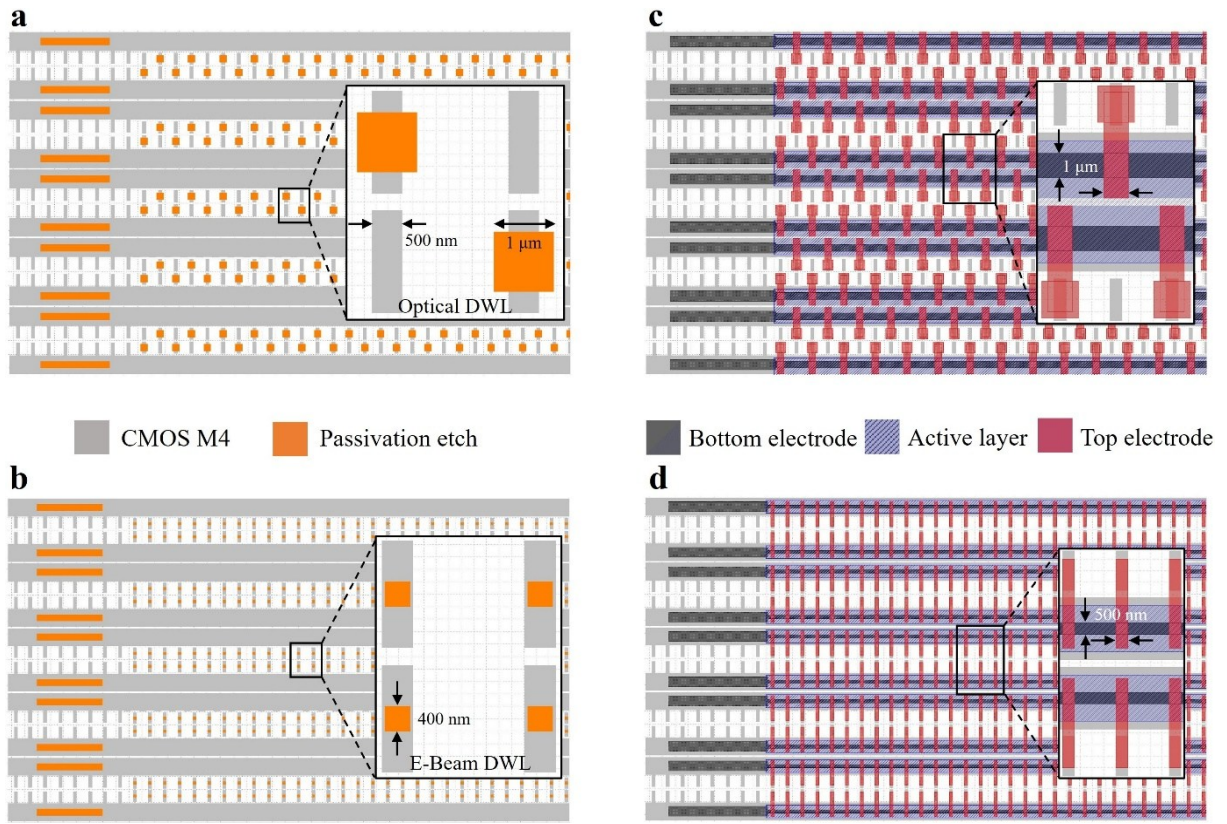

**Supplementary Fig. 2: 1Mbit 1T1R integration layout (virtual masks), designed for optical and e-beam DWL.** Lithographic patterns for a (a) half-array passivation etch (optical), (b) full-array passivation etch (e-beam), (c) half-array RRAM stack (optical) and (d) full-array RRAM stack (e-beam).

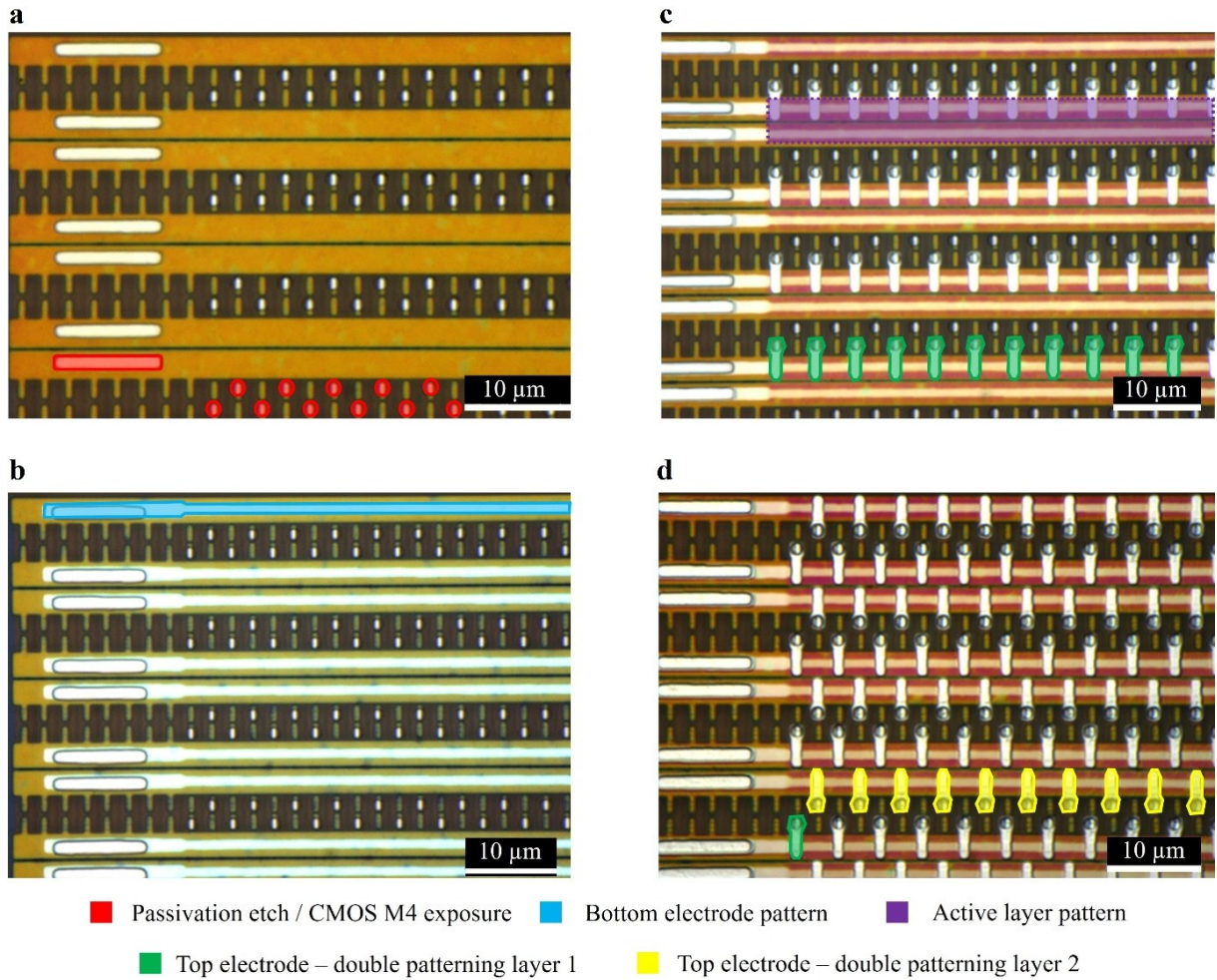

■ Passivation etch / CMOS M4 exposure  ■ Bottom electrode pattern  ■ Active layer pattern
■ Top electrode – double patterning layer 1  ■ Top electrode – double patterning layer 2

**Supplementary Fig. 3: 0.5Mbit 1T1R integration using optical DWL.** (a) Foundry passivation is etched to expose the top CMOS metal. Large openings allow to connect to bottom RRAM electrodes, while small openings allow to connect to top electrodes. (b) Common (per row) bottom electrode patterning. (c) RRAM active layer patterning and patterning of half of the top RRAM electrode features (layer 1). (d) Double patterning employed (layer 2) to fabricate the remaining half of the top RRAM electrode features.

4